\documentclass[preprint,12pt]{elsarticle}

\usepackage{amssymb}
\usepackage{amsmath}
\usepackage{color}
\definecolor{grey}{rgb}{0.8, 0.8, 0.8}
\usepackage{comment}

\biboptions{sort&compress}
\usepackage{cleveref}

\crefname{equation}{}{}
\crefname{figure}{Figure}{Figure}
\crefname{table}{Table}{Table}
\crefname{section}{Section}{Section}

\journal{arXiv}

\begin{document}

\begin{frontmatter}



\title{Non-Bloch band theory for time-modulated discrete mechanical systems}


\author[inst1]{Kei Matsushima}

\affiliation[inst1]{organization={Department of Strategic Studies, Institute of Engineering Innovation, Graduate School of Engineering, The University of Tokyo},
            addressline={2--11--16 Yayoi}, 
            city={Bunkyo--ku},
            postcode={113--8656}, 
            state={Tokyo},
            country={Japan}}

\author[inst1]{Takayuki Yamada}

\begin{abstract}
This study establishes a non-Bloch band theory for time-modulated discrete mechanical systems. We consider simple mass-spring chains whose stiffness is periodically modulated in time. Using the temporal Floquet theory, the system is characterized by linear algebraic equations in terms of Fourier coefficients. This allows us to employ a standard linear eigenvalue analysis. Unlike non-modulated linear systems, the time modulation makes the coefficient matrix non-Hermitian, which gives rise to, for example, parametric resonance, non-reciprocal wave transmission, and non-Hermitian skin effects. In particular, we study finite-length chains consisting of spatially periodic mass-spring units and show that the standard Bloch band theory is not valid for estimating their eigenvalue distribution. To remedy this, we propose a non-Bloch band theory based on a generalized Brillouin zone.
{
    This novel approach, the combination of the temporal Floquet theory for time-modulated systems and generalized Brillouin zone, enables the prediction of eigenvalue distrubution under open boundary conditions and also quantitative characterization of non-Hermitian skin modes.
}
The proposed theory is verified by some numerical experiments.
\end{abstract}



\end{frontmatter}


\section{Introduction}
It is well known that waves propagating in linear and time-independent media obey reciprocity, i.e., transmission between any two points in space is independent of the propagation direction. In periodic lattices, this implies that Bloch bands are symmetric with respect to the inversion of the Bloch wavevector. Reciprocity holds in a number of classical and quantum systems, e.g., photonics, acoustics, and solid mechanics. Recent studies have revealed that reciprocity breaks down in nonlinear or time-varying media \cite{sounas2017non-reciprocal,fleury2015nonreciprocal,nassar2020nonreciprocity}. Broken reciprocity realizes, for example, asymmetric mode conversion \cite{zanjani2014one-way,chen2021efficient}, unidirectional wave propagation \cite{liang2009acoustic,chen2019nonreciprocal,yi2018one-way}, and topologically-protected one-way edge modes \cite{wang2015topological}.

To realize non-reciprocity in mechanical systems, the simplest model would be an array of masses and springs whose constants are modulated in time \cite{nassar2017non-reciprocal,swinteck2015bulk,wang2018observation,wang2015topological,vila2017bloch-based,melkani2023space-time,pu2024multiple}. Without using any nonlinear elements, the time modulation allows us to break time-reversal symmetry and reciprocity of the system. Such time-modulated systems are theoretically tractable with the help of the temporal Floquet theory \cite{ammari2022nonreciprocal,ammari2023transmission,ammari2024topological} and also experimentally feasible using magnets or gyroscopic coupling \cite{wang2015topological,wang2018observation,chen2019nonreciprocal}. 

When a system is periodic both in space and time, the Bloch--Floquet theory formulates a linear or quadratic eigenvalue problem in terms of angular (quasi)frequency $\omega$ for each Bloch wavenumber (wavevector) $k$. When the eigenvalue problem is written in the form $\mathcal{H}(k)\varphi = \omega\varphi$, the square matrix $\mathcal{H}$ (Bloch Hamiltonian) can be non-Hermitian unlike usual time-independent systems. This means that eigenvalues $\omega$ are complex and eigenvectors (eigenmodes) $\varphi$ generally do not form an orthogonal system. Non-Hermiticity gives rise to unique concepts such as parity-time (PT) symmetry \cite{ruter2010observation,matsushima2022unidirectional}, exceptional points \cite{miri2019exceptional,matsushima2023exceptional,even2024experimental}, and non-Hermitian skin effects, which have been intensively studied in the name of non-Hermitian physics \cite{ashida2020non-hermitian}.

The Bloch band theory with spatial periodicity is a useful tool for estimating the spectral properties of finite-size systems. In usual Hermitian systems, it is well known that Bloch bands represent the spectrum of a lattice (chain) whose size (length) is sufficiently large. The non-Hermiticity defies this conventional knowledge. Due to the presence of non-Hermitian skin effects \cite{yao2018edge}, the standard Bloch (quasi-periodic) condition fails to describe some bulk states. This causes significant differences between the spectrum of finite-size systems with open boundary conditions (OBCs) and purely periodic lattices without boundaries, which can be analyzed from the spectral topology in the complex frequency (energy) plane \cite{zhang2020correspondence,zhang2022review}. 

The breakdown of the Bloch band theory motivates us to establish a novel framework that is valid for general non-Hermitian systems. A promising approach is the non-Bloch band theory \cite{kunst2019non-hermitian,yokomizo2019non-bloch} with a generalized Brillouin zone (GBZ) \cite{yao2018edge,yang2020non-hermitian}. The non-Bloch band theory allows the wavenumbers (wavevectors) to be complex such that quasi-periodic eigenmodes represent exponentially growing or decaying bulk states. The non-Bloch band theory is validated in some theoretical models (e.g. Hatano--Nelson model \cite{hatano1996localization} and Su-Schrieffer-Heeger model \cite{su1979solutons}) in quantum physics.

As non-Hermitian mechanical systems are also subjects of intensive study, some theories have been developed and validated. Melkani and Paulose studied time-modulated discrete mechanical systems based on space-time symmetry \cite{melkani2023space-time}. Ghatak et al. discussed the non-Hermitian topology in an active mechanical metamaterial in analogy to the Su-Schrieffer-Heeger model \cite{ghatak2020observation}. A mathematical theory for time-modulated discrete systems is also established by Ammari and Kosche \cite{ammari2024topological}.
{
    Nassar et al. showed that mass-spring chains with space-time modulation break reciprocity \cite{nassar2018quantization}. 
    Salerno et al. studied complex frequency spectra of time-modulated mass-spring lattices and showed that they are associated with parametric instability of mechanical vibration \cite{salerno2016floquet}. 
    Kruss and Paulose discussed non-reciprocal one-way amplification of mechanical waves in time-modulated mass-spring chains using the complex Floquet analysis \cite{kruss2022nondispersive}.
}Nevertheless, it remains unclear how the spectrum of time-modulated dynamic mechanical systems with OBCs is associated with that of periodic lattices.

In this study, we propose a non-Bloch band theory for time-modulated discrete mechanical systems. Time-modulated mass-spring chains are formulated using the temporal Floquet theory.
{
    Our model is similar to the one analyzed in \cite{nassar2018quantization}; however, we generalize their Bloch--Floquet theory to complex frequency spectra for the analysis of the GBZ and non-Hermitian skin effects.
}We present numerical evidence that the conventional Bloch band theory is not valid for the time-modulated systems. Following \cite{yokomizo2019non-bloch}, we define a {GBZ} and establish a non-Bloch band theory. This allows us to estimate eigenvalues of long chains in the complex frequency plane even in the presence of non-Hermitian skin modes that arise from the non-reciprocity. 

\section{Chain of a finite number of masses and springs}
\subsection{Model}
\begin{figure}
    \centering
    \includegraphics[scale=0.8]{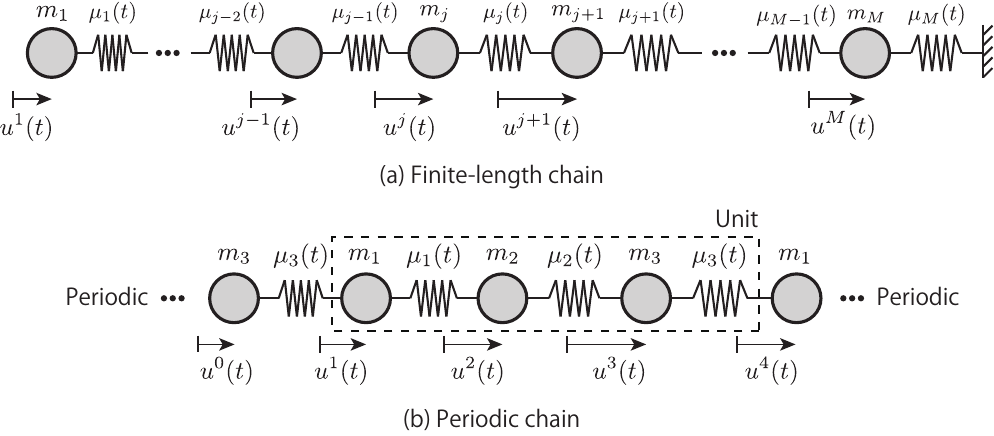}
    \caption{Chain of spring-mass pairs. }
    \label{fig:chain-finite}
\end{figure}
As shown in Figure \ref{fig:chain-finite} (a), we consider a chain of masses $m_1,m_2,\ldots,m_M$ connected with springs whose parameters are $\mu_1,\mu_2,\ldots,\mu_{M}$. The parameters (stiffness) $\mu_1,\ldots,\mu_{M}$ are modulated in time while the masses are constant. The displacement $u^j$ of each mass ($j=1,\ldots,M)$ is governed by the following equations of motion:
\begin{align}
    m_j \partial_{tt} u^j(t) = \mu_{j-1}(t) u^{j-1}(t) - (\mu_{j-1}(t)+\mu_j(t)) u^j(t) + \mu_j u^{j+1}(t) + f_j(t) \label{eq:tmp1}
\end{align}
for $j=2,\ldots,M-1$, and
\begin{align}
    m_1 \partial_{tt} u^1(t) = - \mu_1(t) u^1(t) + \mu_1 u^{2}(t) + f_1(t), \label{eq:tmp2}
    \\
    m_M \partial_{tt} u^M(t) = \mu_{M-1}(t) u^{M-1}(t) - (\mu_{M-1}(t)+\mu_M) u^M(t) + f_M(t), \label{eq:tmp3}
\end{align}
where $f_j$ denotes an external force applied to the mass $m_j$.

Let us consider a time-harmonic excitation, i.e., for each $j=1,\ldots,M$, there exists a constant $F_j\in\mathbb{C}$ such that
\begin{align}
    f_j(t) = \mathrm{Re}\left[ F_j \mathrm{e}^\mathrm{-\mathrm{i}\omega t} \right] \label{eq:tmp4}
\end{align}
with angular frequency $\omega\in\mathbb{R}$. In addition, we assume that the modulation of the stiffness is periodic in time with angular frequency $\Omega>0$ and given by
\begin{align}
    \mu_j(t) = G_j + \gamma_j\cos(\Omega t - \phi_j) = G_j + \frac{\gamma_j}{2} \mathrm{e}^{-\mathrm{i} \phi_j} \mathrm{e}^{\mathrm{i}\Omega t} + \frac{\gamma_j}{2} \mathrm{e}^{\mathrm{i} \phi_j} \mathrm{e}^{-\mathrm{i}\Omega t} \label{eq:tmp5}
\end{align}
with constants $G_j>0$, $\gamma_j\in [0,G_j)$, and $\phi_j\in[-\pi,\pi]$. Since all the coefficients $\mu_j$ are $T$-periodic, i.e., $\mu_j(t+T) = \mu_j(t)$ with $T=2\pi/\Omega$ for all $j=1,\ldots,M$ and $t\in\mathbb{R}$, we use the Floquet theory \cite{ammari2022nonreciprocal,ammari2023transmission,ammari2024topological} and seek solutions $u^j$ of the form
\begin{align}
    u^j(t) =& \mathrm{Re}\left[ \mathrm{e}^{-\mathrm{i}\omega t} \sum_{n=-\infty}^\infty U^j_n \mathrm{e}^{-\mathrm i n\Omega t} \right] \label{eq:tmp6},
\\
    \partial_t u^j(t) =& \mathrm{Re}\left[ \mathrm{e}^{-\mathrm{i}\omega t} \sum_{n=-\infty}^\infty V^j_n \mathrm{e}^{-\mathrm i n\Omega t} \right]
\end{align}
with unknown coefficients $U^j_n,V^j_n\in\mathbb{C}$ ($n\in\mathbb{Z}$) for each $j=1,\ldots,M$. The coefficients should satisfy $(\omega+n\Omega) U^j_n = \mathrm{i} V^j_n$ 
for each $j$ and $n$. Note that the response $u^j$ is not in general time-harmonic with angular frequency $\omega$ despite the harmonic excitation $f_j$. In view of this, the factor $\omega$ in \cref{eq:tmp6} is called \textit{quasifrequency} \cite{ammari2023transmission}. 

Inserting \cref{eq:tmp4,eq:tmp5,eq:tmp6} into \cref{eq:tmp1,eq:tmp2,eq:tmp3}, we obtain the following linear relation for each $n\in\mathbb{Z}$:
\begin{align}
    &- \mathrm{i} m_j (\omega + n\Omega) V^j_n -G_{j-1} U^{j-1}_n + (G_{j-1} + G_j  ) U^j_n - G_j U^{j+1}_n 
\notag\\
    & -\bar\eta_{j-1} U^{j-1}_{n+1} + (\bar\eta_{j-1} + \bar\eta_j) U^j_{n+1} - \bar\eta_j U^{j+1}_{n+1}
\notag\\
    & -\eta_{j-1} U^{j-1}_{n-1} + (\eta_{j-1} + \eta_j) U^j_{n-1} - \eta_j U^{j+1}_{n-1} = F_j\delta_{n,0}, \quad j=2,\ldots,M-1, \label{eq:linear-sys_1}
\\
    & - \mathrm{i}m_1 (\omega + n\Omega)V^1_n  + G_1  U^1_n - G_1 U^{2}_n
\notag\\
    &\qquad + \bar\eta_1 U^1_{n+1} - \bar\eta_1 U^{2}_{n+1} + \eta_1 U^1_{n-1} - \eta_j U^{2}_{n-1} = F_1\delta_{n,0}, \label{eq:linear-sys_2}
\\
    &-\mathrm{i} m_M (\omega + n\Omega)V^M_n -G_{M-1} U^{M-1}_n + ( G_{M-1} + G_M  ) U^M_n 
\notag\\
    &\qquad -\bar\eta_{M-1} U^{M-1}_{n+1} + (\bar\eta_{M-1} + \bar\eta_M) U^M_{n+1} 
\notag\\
    &\qquad -\eta_{M-1} U^{M-1}_{n-1} + (\eta_{M-1} + \eta_M) U^M_{n-1}  = F_M\delta_{n,0}, \label{eq:linear-sys_3}
\end{align}
where $\eta_j$ is the constant defined by $\eta_j = \gamma_j \mathrm{e}^{\mathrm i\phi_j}/2$, and $\delta_{n,0}$ is the Kronecker delta, i.e.,
\begin{align}
    \delta_{n,0} = 
    \begin{cases}
        1 & (n=0) \\ 0 & (n\neq 0)
    \end{cases}
    .
\end{align}
For numerical computation, we truncate the Fourier series in terms of $n$ with sufficiently large integer $P$ and approximate the linear system \cref{eq:linear-sys_1,eq:linear-sys_2,eq:linear-sys_3} into
\begin{align}
    \left(
     \begin{bmatrix}
         \mathbf{K} & -\mathrm{i}\Omega \mathbf{M} \mathbf{B}
         \\
         \mathrm{i}\Omega \mathbf{M} \mathbf{B} & \mathbf{M}
     \end{bmatrix}
     - \omega
     \begin{bmatrix}
         O & \mathrm{i}\mathbf{M}
         \\
         -\mathrm{i}\mathbf{M} & O
     \end{bmatrix}
    \right)
    \begin{pmatrix}
        U \\ V
    \end{pmatrix}
    =
    \begin{pmatrix}
        F \\ 0
    \end{pmatrix}
    ,
    \label{eq:eq-finite}
\end{align}
where the matrices are defined by
\begin{align}
    \mathbf{K} =& 
    \begin{bmatrix}
         K_1 & -K_1    &         &           &
        \\
        -K_1 & K_1+K_2 & -K_2    &           &
        \\
             & -K_2    & K_2+K_3 & -K_3      &
        \\
             &         &         & \ddots    &
        \\
             &         &         &  -K_{M-1} & K_{M-1}+K_M
    \end{bmatrix},
\\
    \mathbf{M} =& \mathrm{diag}(m_1 I, m_2 I, \cdots, m_M I),
\\
    \mathbf{B} =& \mathrm{diag}(B, B, \cdots, B)
\end{align}
with submatrices
\begin{align}
    K_j = 
    \begin{bmatrix}
        G_j    & \bar\eta_j &            &            &
        \\
        \eta_j & G_j        & \bar\eta_j &            &
        \\
               & \eta_j     & G_j        & \bar\eta_j &
        \\
               &            &            & \ddots     &
        \\         
               &            &            & \eta_j     & G_j
    \end{bmatrix}
    , \quad
    B = \mathrm{diag}(-P,-P+1,\ldots,P-1,P).
\end{align}
Analogously, the vectors are associated with the constants and Fourier coefficients as follows:
\begin{align}
    U =& (U^1_{-P},\ldots,U^1_0,\ldots,U^1_{P}, \ldots, U^M_{-P},\ldots,U^M_0,\ldots,U^M_{P})^T,
\\
    V =& (V^1_{-P},\ldots,V^1_0,\ldots,V^1_{P}, \ldots, V^M_{-P},\ldots,V^M_0,\ldots,V^M_{P})^T,
\\
    F =& (0,\ldots,F_1,\ldots,0, \ldots, 0,\ldots,{F_M},\ldots,0)^T.
\end{align}
Note that the matrix product
\begin{align}
\mathcal{H} := 
\begin{bmatrix}
         O & \mathrm{i}\mathbf{M}
         \\
         -\mathrm{i}\mathbf{M} & O
     \end{bmatrix}^{-1}
    \begin{bmatrix}
         \mathbf{K} & -\mathrm{i}\Omega \mathbf{M} \mathbf{B}
         \\
         \mathrm{i}\Omega \mathbf{M} \mathbf{B} & \mathbf{M}
     \end{bmatrix}
\end{align}
is not necessarily Hermitian despite that each matrix is Hermitian. Thus, we cannot always expect that eigenvalues $\omega$ of $\mathcal{H}$ are all real. However, due to the Hermiticity of the two matrices, it can be proved that the eigenvalues $\omega$ are real or come in complex-conjugate pairs $(\omega,\bar\omega)$ \cite{tisseur2001quadratic}.

\subsection{Numerical example}\label{ss:example}
\begin{figure}
    \centering
    \includegraphics[scale=0.79]{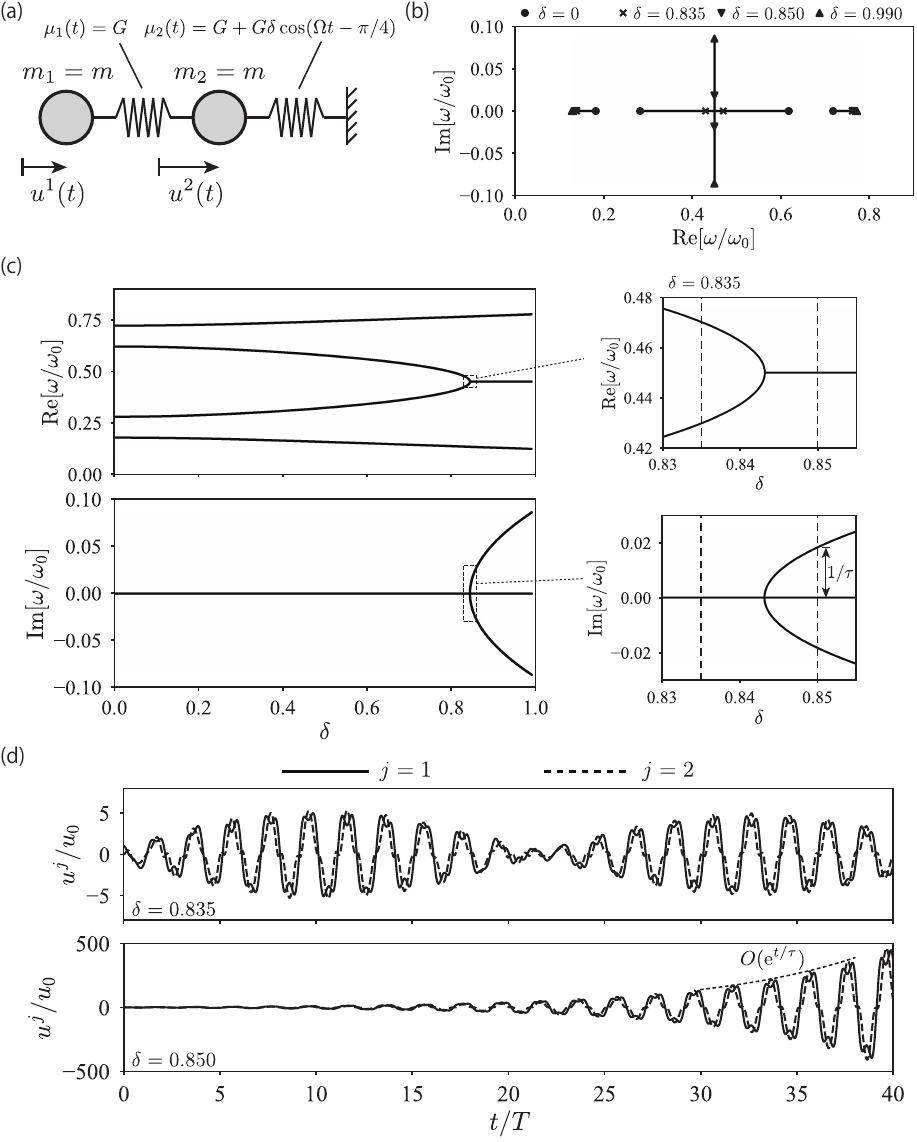}
    \caption{Verification of the Floquet theory for a time-modulated system. (a) Two mass-spring pairs. (b), (c) Trajectory of eigenvalues $\omega$ as the modulation amplitude $\delta$ varies from $0$ to $0.99$. (d) {D}isplacements $u^1$ and $u^2$ as a function of time $t$ when the initial conditions $u^1(0)=u_0$ and $u^2(0)=\partial_t u^1(0) = \partial_t u^2(0)=0$ are given.}
    \label{fig:result-verification_result}
\end{figure}

We first check that the Floquet theory is valid for the time-modulated system. As shown in \cref{fig:result-verification_result} (a), we consider the mass-spring pairs with $M=2$. The masses are set to $m_1 = m_1 = m$ with given $m>0$, and the stiffness is given by $\mu_1(t) = G$ and $\mu_2(t) = G + G\delta\cos(\Omega t-\pi/4)$ for some $G>0$ and $\delta \in [0,1)$ (i.e., $\gamma_1=0$, $\gamma_2 = G\delta$, and $\phi_2=\pi/4$). The modulation angular frequency is $\Omega = 0.9\omega_0$, where $\omega_0 = \sqrt{G/m}$ is the resonant frequency of the single mass-spring pair. For numerical computation, the Fourier series in terms of $n\in\mathbb{Z}$ are truncated into $n=-4,\ldots,4$, i.e., $P=4$. 

\cref{fig:result-verification_result} (b) and (c) show the trajectories of eigenvalues $\omega$ of $\mathcal{H}$ as the modulation amplitude $\delta$ varies from $0$ to $0.99$. As the spectrum is $\Omega$-periodic in the complex $\omega$-plane \cite{ammari2023transmission}, we plot eigenvalues only within the range of $0\leq \mathrm{Re}[\omega] \leq \Omega$. The results indicate that all the eigenvalues are purely real when the modulation amplitude $\delta$ is less than $0.843$. Above this threshold, the eigenvalues come in a conjugate pair $(\omega,\bar\omega)$ with nonzero imaginary part. The eigenvalues with negative and positive imaginary part are associated with modes decaying and growing in time, respectively (recall that the convention $\mathrm{e}^{-\mathrm i\omega t}$ is used). This means that the system undergoes a phase transition from stable to unstable states. This is a typical phenomenon in PT symmetric systems and called real-to-complex eigenvalue transition, pseudo-Hermiticity breaking, or PT phase transition \cite{mostafazadeh2002pseudo-hermiticity,melkani2023space-time}.

We compare the results of the eigenvalue analysis with numerical solutions of \cref{eq:tmp1,eq:tmp2,eq:tmp3}. As \cref{eq:tmp1,eq:tmp2,eq:tmp3} are just a linear system of ordinary differential equations, we use the backward Euler method to stably compute its numerical solution. The temporal discretization is given by $t=0,\Delta t, 2\Delta t, \ldots$ with $\Delta t= T / 5000$. \cref{fig:result-verification_result} (d) shows numerical solutions $u^1(t)$ and $u^2(t)$ when the initial conditions are given by $u^1(0) = u_0$ and $u^2(0)=\partial_t u^1(0) = \partial_t u^2(0) = 0$ for some $u_0>0$. For $\delta=0.835$, it is clear that the oscillation is stable in time. In the case of $\delta = 0.850$, the displacements grow exponentially in time. This is consistent with the discussion from the Floquet theory. Moreover, their growth rate $1/\tau$ is almost the same with the imaginary part of the eigenvalue $\omega$ obtained from the Floquet theory (see \cref{fig:result-verification_result} (c)). From these results, we conclude that the Floquet theory is valid for the time-modulated mechanical system.

\begin{figure}
    \centering
    \includegraphics[scale=0.76]{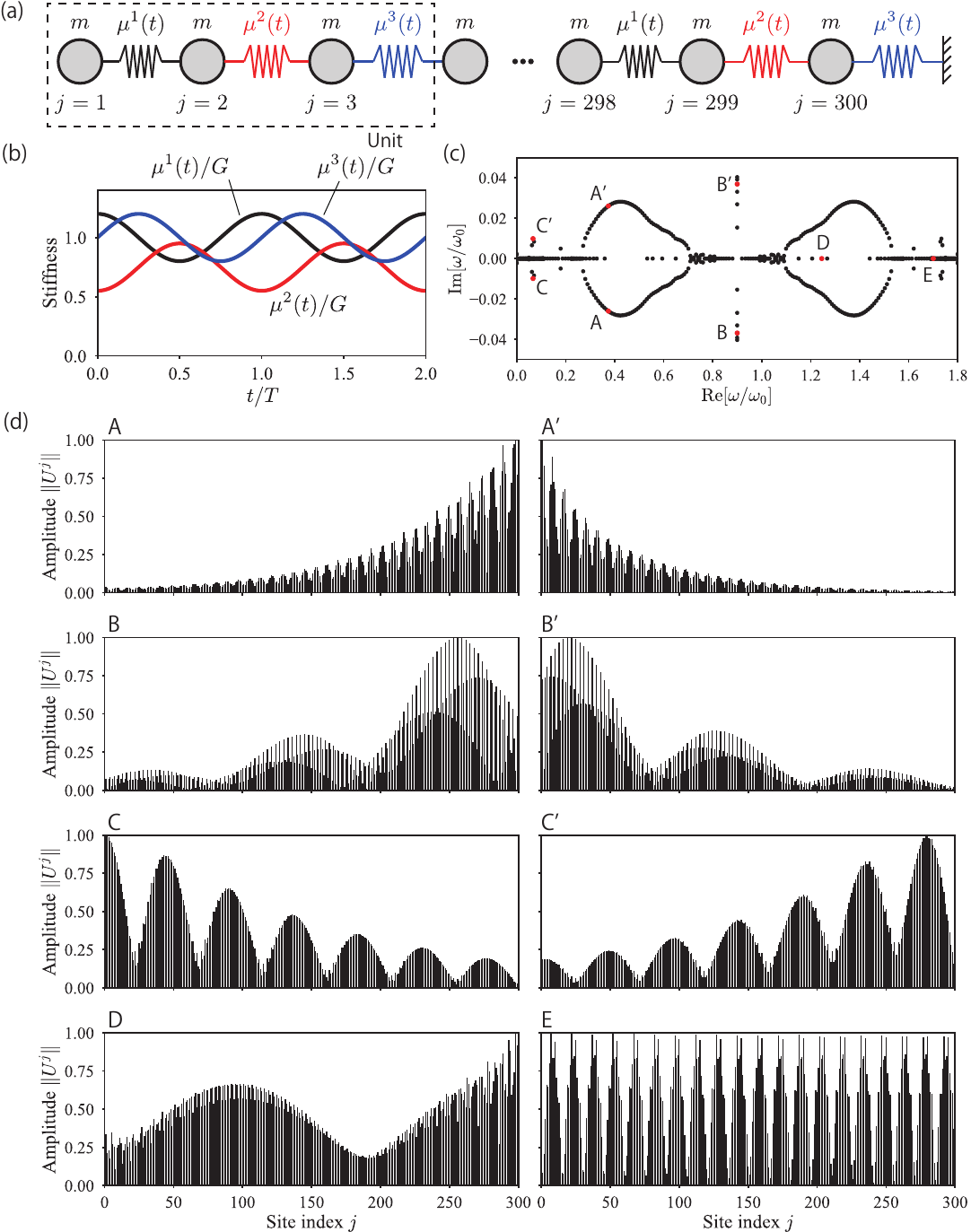}
    \caption{Spectrum of $\mathcal{H}$ for a long-chain of mass-spring pairs. (a) A long chain consists of spatially periodic units. Its unit comprises three mass-spring pairs. (b) Time modulation of {$\mu_1$, $\mu_2$, and $\mu_3$}. (c) Distribution of eigenvalues $\omega$ for the system (a). Some representatives are denoted by the symbols A, B, C, D, and E with their conjugate pairs. (d) Eigenmodes corresponding to the representatives.}
    \label{fig:result-finite-spectrum_result}
\end{figure}

We are interested in the spectrum of time-modulated chains with large $M$. As shown in \cref{fig:result-finite-spectrum_result} (a) and (b), we consider the chain of length $M=300$ with the following parameters: 
\begin{align}
    \Omega = 1.8\omega_0 ,\quad m_1 = \ldots = m_{300} = m, \quad  \gamma_1 = \ldots = \gamma_{300} = 0.2G, \label{eq:param-1}
\\
    G_{3j-2} = G , \quad G_{3j-1} = 0.75G, \quad G_{3j} = G \quad (j=1,\ldots,100), \label{eq:param-2}
\\
    \phi_{3j-2} = 0 , \quad \phi_{3j-1} = \pi, \quad \phi_{3j} = \pi/2 \quad (j=1,\ldots,100). \label{eq:param-3}
\end{align}

\cref{fig:result-finite-spectrum_result} (c) shows the distribution of the eigenvalues $\omega$ of $\mathcal{H}$. In analogy with the previous results with $M=2$, the spectrum is symmetric with respect to the real axis. In addition, the spectrum accumulates along some curves in the complex $\omega$-plane. 

In \cref{fig:result-finite-spectrum_result} (d), we plot eigenmodes corresponding to some eigenvalues in the spectrum. The eigenmodes A, B, and C (and their conjugate pairs) exponentially grow as the site index $j$ increases or decreases. This means that some modes are localized at either left or right edge of the chain. The other modes D and E (purely real eigenvalues) are almost uniformly distributed in space. This localization deserves a careful investigation as the usual mass-spring chains do not exhibit such spatial behaviors.

\section{Non-Bloch band theory}
In the previous section, we observed that the long chain with repeating units exhibit some special spectral structures and its eigenmodes can be localized at the edges. To analyze these unique properties, we approximate the long chain as a purely periodic system. The interesting point is that the standard Bloch band theory breaks down for our time-modulated system due to the non-Hermiticity of $\mathcal{H}$. In this section, we introduce the non-Bloch band theory \cite{kunst2019non-hermitian,yokomizo2019non-bloch} with a {GBZ} \cite{yao2018edge,yang2020non-hermitian} to study the spectrum of $\mathcal{H}$ for long chains with OBCs.

\subsection{Bloch band theory with Brillouin zone}\label{ss:bloch}
As shown in \cref{fig:chain-finite} (b), we consider a periodic chain of masses and springs. The unit of the chain comprises three masses $(m_1,m_2,m_3)$ connected with springs characterized by $(\mu_1,\mu_2,\mu_3)$. As in the previous section, the masses are constant in time, and the stiffness $\mu_j$ is modulated as \cref{eq:tmp5} using constants $G_j$, $\gamma_j$, and $\phi_j$ for $j=1,2,3$. Since we are interested in the spectrum, the external force is set to $f_j=0$. For each $j\in\mathbb{Z}$, the displacement $u^j$ satisfies the equation of motion \cref{eq:tmp1}. The periodicity with respect to time $t$ allows us to use the Floquet condition \cref{eq:tmp6}. In addition, since the system is now also periodic in space, the Bloch theory states that a solution is quasi-periodic in space, i.e., there exists a scalar $k$ (Bloch wavenumber) such that
\begin{align}
    U^{j+3}_n = U^j_n \mathrm{e}^{\mathrm i k} \label{eq:bloch}
\end{align}
for all $j\in\mathbb{Z}$. Then, we arrive at the following linear system
\begin{align}
    \left(
     \begin{bmatrix}
         \mathbf{K}_k & -\mathrm{i}\Omega \mathbf{M} \mathbf{B}
         \\
         \mathrm{i}\Omega \mathbf{M} \mathbf{B} & \mathbf{M}
     \end{bmatrix}
     - \omega
     \begin{bmatrix}
         O & \mathrm{i}\mathbf{M}
         \\
         -\mathrm{i}\mathbf{M} & O
     \end{bmatrix}
    \right)
    \begin{pmatrix}
        U \\ V
    \end{pmatrix}
    =
    \begin{pmatrix}
        0 \\ 0
    \end{pmatrix}
    ,
    \label{eq:eq-periodic}
\end{align}
where the matrix $\mathbf{K}_k$ is defined by
\begin{align}
    \mathbf{K}_k = 
    \begin{bmatrix}
         K_3 + K_1 & -K_1    &   -K_3\mathrm{e}^{-\mathrm ik}
        \\
        -K_1 & K_1+K_2 & -K_2    
        \\
        -K_3\mathrm{e}^{\mathrm ik}     & -K_2    & K_2+K_3 
    \end{bmatrix}
    ,
\end{align}
and the other matrices and vectors are as in the previous section with $M=3$.

For fixed $k$, the system \cref{eq:eq-periodic} can be rewritten as $\mathcal{H}_k\varphi = \omega\varphi$ with vector $\varphi$ and square matrix (Bloch Hamiltonian) 
\begin{align}
\mathcal{H}_k := 
    \begin{bmatrix}
         O & \mathrm{i}\mathbf{M}
         \\
         -\mathrm{i}\mathbf{M} & O
     \end{bmatrix}^{-1}
     \begin{bmatrix}
         \mathbf{K}_k & -\mathrm{i}\Omega \mathbf{M} \mathbf{B}
         \\
         \mathrm{i}\Omega \mathbf{M} \mathbf{B} & \mathbf{M}
     \end{bmatrix}
,    
\end{align}
which depends on $k$. Let $\sigma(k) = \{ \omega\in\mathbb{C} :  \mathrm{det}(\omega I-\mathcal{H}_k) = 0 \}$ be the spectrum of $\mathcal{H}_k$. 
The union $\sigma_\mathrm{BZ}:= \cup_{k\in\mathbb R}\ \sigma(k) $ is called a Bloch band. For the computation of $\sigma_\mathrm{BZ}$, it suffices to consider $k\in[-\pi,\pi]$ as the $2\pi$ shift does not affect the Bloch condition \cref{eq:bloch}, i.e., $\mathcal{H}_{k+2\pi} = \mathcal{H}_{k}$ and $\sigma_\mathrm{BZ}= \cup_{k\in[-\pi,\pi]}\ \sigma(k) $. The range $[-\pi,\pi]$ is called the (first) Brillouin zone (BZ). 

For Hermitian systems (e.g. $\gamma_j=0$), it is well known that the Bloch band $\sigma_\mathrm{BZ}$ is an approximation to the spectrum $\sigma_\mathrm{OBC}:= \{ \omega\in\mathbb{C} :  \mathrm{det}(\omega I-\mathcal{H}) = 0 \}$ of the long chain (sufficiently large $M$) with OBCs, discussed in the previous section. Namely, the OBCs can be regarded as a small perturbation to the purely periodic system without edges.

However, this well-known fact does not always hold true for non-Hermitian systems \cite{zhang2022review}. The Bloch condition \cref{eq:bloch} with real $k$ implicitly enforces solutions not to grow or decay in space, which is not a reasonable assumption from the results in \cref{fig:result-finite-spectrum_result} (d).

\begin{figure}
    \centering
    \includegraphics[scale=0.75]{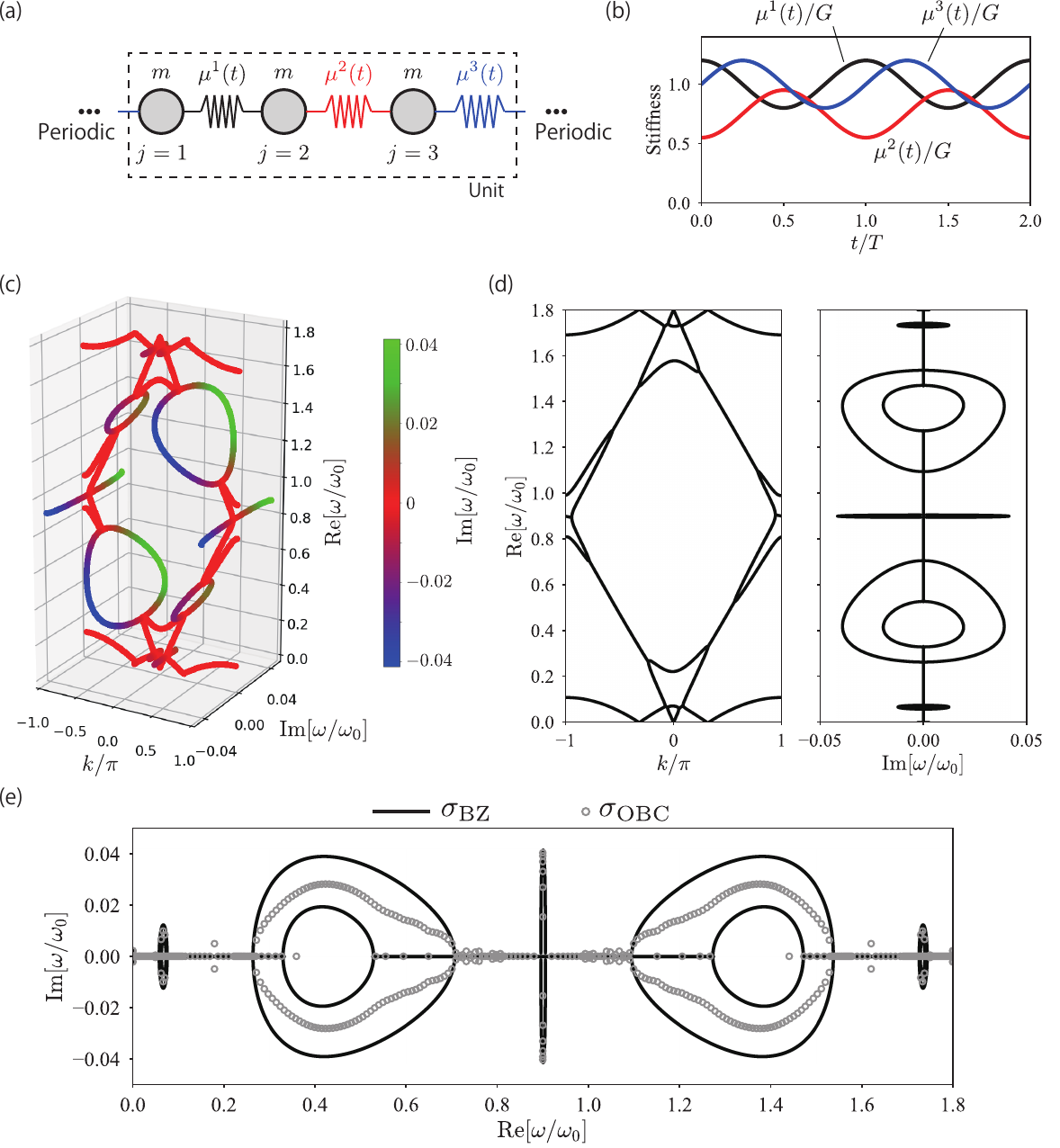}
    \caption{Bloch band $\sigma_\mathrm{BZ}$ and OBC spectrum $\sigma_\mathrm{OBZ}$. (a), (b) Periodic chain of mass-spring pairs. The unit of the chain is the same as in \cref{fig:result-finite-spectrum_result} (a), (b). (c), (d) $\sigma(k)$ as a function of $k\in[-\pi,\pi]$ (Bloch band diagram). (e) Comparison between $\sigma_\mathrm{BZ}$ and $\sigma_\mathrm{OBZ}$ in \cref{fig:result-finite-spectrum_result}.}
    \label{fig:result-periodic-spectrum_result}
\end{figure}

To see this, we present a numerical example of the Bloch band $\sigma_\mathrm{BZ}$ in \cref{fig:result-periodic-spectrum_result}. As shown in \cref{fig:result-periodic-spectrum_result} (a) and (b), we use the same modulation frequency $\Omega$ and material parameters $m_j$, $G_j$, $\gamma_j$, and $\phi_j$ in \cref{eq:param-1,eq:param-2,eq:param-3}. \cref{fig:result-periodic-spectrum_result} (c) and (d) show the Bloch band diagram on the {BZ} $-\pi\leq k \leq \pi$. As the periodic system is also non-Hermitian, the eigenvalues of $\mathcal{H}_k$ are complex-valued in general. Notably, the bands are asymmetric with respect to the inversion $k\to -k$, which implies that the system is non-reciprocal. \cref{fig:result-periodic-spectrum_result} (e) shows the comparison between the bands $\sigma_\mathrm{BZ}$ and spectrum of the long chain $\sigma_\mathrm{OBC}$ in \cref{fig:result-finite-spectrum_result} (c). From the result, we observe the obvious discrepancy between the two spectra. This implies that the Bloch band theory with the standard BZ does not work for our time-modulated system.

\subsection{Non-Bloch band theory with generalized Brillouin zone}
To deal with eigenmodes growing and decaying in space, we need to allow the Bloch wavenumber $k$ to be complex. Following the common notation, we define $\beta = \mathrm{e}^{\mathrm ik}$. Then, the linear system \cref{eq:eq-periodic} is rewritten as
\begin{align}
    \left(
     \begin{bmatrix}
         \mathbf{P}_0 - \beta \mathbf{P}_1 - \beta^{-1} \mathbf{P}_{-1} & -\mathrm{i}\Omega \mathbf{M} \mathbf{B}
         \\
         \mathrm{i}\Omega \mathbf{M} \mathbf{B} & \mathbf{M}
     \end{bmatrix}
     - \omega
     \begin{bmatrix}
         O & \mathrm{i}\mathbf{M}
         \\
         -\mathrm{i}\mathbf{M} & O
     \end{bmatrix}
    \right)
    \begin{pmatrix}
        U \\ V
    \end{pmatrix}
    =
    \begin{pmatrix}
        0 \\ 0
    \end{pmatrix}
    ,
    \label{eq:periodic-beta}
\end{align}
where the matrices $\mathbf P_0$, $\mathbf P_1$, and $\mathbf P_{-1}$ are defined by
\begin{align}
    \mathbf{P}_0 &= 
    \begin{bmatrix}
         K_3 + K_1 & -K_1    &   O
        \\
        -K_1 & K_1+K_2 & -K_2    
        \\
        O     & -K_2    & K_2+K_3 
    \end{bmatrix}
    ,
\\
    \mathbf{P}_1 &= 
    \begin{bmatrix}
       O  & O & O 
        \\
       O & O &     O
        \\
        K_3     &   O  & O
    \end{bmatrix}
    ,
\quad
    \mathbf{P}_{-1} = 
    \begin{bmatrix}
       O  & O & K_3
        \\
       O & O &     O
        \\
       O    &   O  & O
    \end{bmatrix}
    .
\end{align}
The system \cref{eq:periodic-beta} yields a (generalized) linear eigenvalue problem in terms of $\omega$ for fixed $\beta\in\mathbb{C}$. When $\omega\in\mathbb{C}$ is given, this is a quadratic eigenvalue problem in terms of $\beta$, which can be recast into a linear eigenvalue problem \cite{tisseur2001quadratic}.

For each $\omega\in\mathbb{C}$, let $\beta_1(\omega),\beta_2(\omega),\ldots,\beta_{2 N}(\omega)$ be eigenvalues of \cref{eq:periodic-beta}, where {$N:=6(2P+1)$ is the size of the whole matrix in \cref{eq:periodic-beta}}. The eigenvalues are sorted such that $|\beta_1(\omega)|\leq |\beta_2(\omega)| \leq \ldots \leq |\beta_{2N}(\omega)|$.
A generalized Brillouin zone (GBZ) \cite{yao2018edge,yang2020non-hermitian} is defined by
\begin{align}
    C_\mathrm{GBZ} = \{ \beta_{N}(\omega) : \omega\in\sigma_\mathrm{GBZ} \} \cup \{ \beta_{N+1}(\omega) : \omega\in\sigma_\mathrm{GBZ} \}, \label{eq:GBZ}
\end{align}
with
\begin{align}
    \sigma_\mathrm{GBZ} = \{ \omega\in\mathbb{C} : |\beta_{ N}(\omega)| = |\beta_{ N+1}(\omega)| \}.
\end{align}
The non-Bloch band theory claims that the set $\sigma_\mathrm{GBZ}$ is an approximation to the spectrum of $\mathcal{H}$ with repeating units and large $M$ (e.g. \cref{fig:result-finite-spectrum_result} (a)).

{
    To see this, let $(\tilde U^1_l,\tilde U^2_l,\tilde U^3_l,\tilde V^1_l,\tilde V^2_l,\tilde V^3_l)$ be an eigenvector of \cref{eq:periodic-beta} corresponding to $\beta_l$ for each $l=1,\ldots,2N$ (here we omit the dependence on $\omega$). Let us consider the finite-length system \cref{eq:eq-finite} with $F=0$, $M=3L$, and repeating constants, e.g., $m_1=m_4=\ldots=m_{3L-2}$, $m_2=m_5=\ldots=m_{3L-1}$, and $m_3=m_6=\ldots=m_{3L}$, where $L$ is the number of units. We assume that a solution of \cref{eq:eq-finite} can be written as
    \begin{align}
        \begin{pmatrix}
            U^{3j-2} \\ U^{3j-1} \\ U^{3j} \\ 
            V^{3j-2} \\ V^{3j-1} \\ V^{3j} 
        \end{pmatrix}
        =
        \sum_{l=1}^{2N} c_l \beta_l^{j-1} 
        \begin{pmatrix}
            \tilde U^{1}_l \\ \tilde U^{2}_l \\ 
            \tilde U^{3}_l \\ \tilde V^{1}_l \\ 
            \tilde V^{2}_l \\ \tilde V^{3}_l 
        \end{pmatrix}
    \end{align}
    for all $j=1,\ldots,L$. The coefficients $c_1,\ldots,c_{2N}\in\mathbb C$ are determined from the equations of motion and OBCs. In what follows, we limit ourselves to the homogeneous Dirichlet condition $(U^1,V^1) = (U^{3L},V^{3L}) = (0,0)$. Other linear and homogeneous OBCs can be treated analogously.

    From the OBC at the left end, we have
    \begin{align}
        \begin{pmatrix}
            0 \\ 0
        \end{pmatrix}
        =
        \begin{pmatrix}
            U^1 \\ V^1
        \end{pmatrix}
        =
        \sum_{l=1}^{2N} c_l 
        \begin{pmatrix}
            \tilde U_l^1 \\ \tilde V_l^1
        \end{pmatrix}
        ,
    \end{align}
    or equivalently
    \begin{align}
        \begin{bmatrix}
            \varphi^1_1 & \varphi^1_2 & \cdots & \varphi^1_{2N}
        \end{bmatrix}
        \begin{pmatrix}
            c_1 \\ c_2 \\ \vdots \\ c_{2N}
        \end{pmatrix}
        = 0
    \end{align}
    with $\varphi^1_l := (\tilde U_l^1, \tilde V_l^1)\in \mathbb C^{N/3}$ for each $l=1,\ldots,2N$. Analogously, the OBC at the right end yields
    \begin{align}
        \begin{bmatrix}
            \beta_1^{L-1} \psi^1_1 & \beta_2^{L-1} \psi^1_2 & \cdots & \beta_{2N}^{L-1} \psi^1_{2N}
        \end{bmatrix}
        \begin{pmatrix}
            c_1 \\ c_2 \\ \vdots \\ c_{2N}
        \end{pmatrix}
        = 0,
    \end{align}
    where $\psi^1_1,\ldots,\psi^1_{2N}\in \mathbb C^{N/3}$ are defined by $\psi^1_l := (\tilde U_l^3, \tilde V_l^3)$.

    From the equation of motion for the leftmost mass (the first row of \cref{eq:eq-finite}), we see that $(U^1,V^1)$ and $(U^2,V^2)$ satisfy the following condition:
    \begin{align}
        \begin{bmatrix}
            K_1 & -\mathrm im_1(\Omega B + \omega I)
            \\
            \mathrm im_1(\Omega B + \omega I) & m_1 I
        \end{bmatrix}
        \begin{pmatrix}
            U^1 \\ V^1
        \end{pmatrix}
        +
        \begin{bmatrix}
            -K_1 & O
            \\
            O & O
        \end{bmatrix}
        \begin{pmatrix}
            U^2 \\ V^2
        \end{pmatrix}
        =
        \begin{pmatrix}
            0 \\ 0
        \end{pmatrix}
        ,
    \end{align}
    which can be rewritten as
    \begin{align}
        \begin{bmatrix}
            \varphi^2_1 & \varphi^2_2 & \cdots & \varphi^2_{2N}
        \end{bmatrix}
        \begin{pmatrix}
            c_1 \\ c_2 \\ \vdots \\ c_{2N}
        \end{pmatrix}
        = 0,
    \end{align}
    where $\varphi^2_1,\ldots,\varphi^2_{2N}\in \mathbb C^{N/3}$ are defined by
    \begin{align}
        \varphi^2_l := 
        \begin{bmatrix}
            K_1 & -\mathrm im_1(\Omega B + \omega I)
            \\
            \mathrm im_1(\Omega B + \omega I) & m_1 I
        \end{bmatrix}
        \begin{pmatrix}
            \tilde U_l^1 \\ \tilde V_l^1
        \end{pmatrix} 
        + 
        \begin{bmatrix}
            -K_1 & O
            \\
            O & O
        \end{bmatrix}
        \begin{pmatrix}
            \tilde U_l^2 \\ \tilde V_l^2
        \end{pmatrix} 
    \end{align}
    for all $l=1,\ldots,2N$.

    The same argument can be applied to the second and last two masses. We finally arrive at the following $2N\times 2N$ linear system:
    \begin{align}
        \begin{bmatrix}
            \varphi^1_1 & \varphi^1_2 & \cdots & \varphi^1_{2N}
            \\
            \varphi^2_1 & \varphi^2_2 & \cdots & \varphi^2_{2N}
            \\
            \varphi^3_1 & \varphi^3_2 & \cdots & \varphi^3_{2N}
            \\
            \beta_1^{L-1} \psi^1_1 & \beta_2^{L-1} \psi^1_2 & \cdots & \beta_{2N}^{L-1} \psi^1_{2N}
            \\
            \beta_1^{L-1} \psi^2_1 & \beta_2^{L-1} \psi^2_2 & \cdots & \beta_{2N}^{L-1} \psi^2_{2N}
            \\
            \beta_1^{L-1} \psi^3_1 & \beta_2^{L-1} \psi^3_2 & \cdots & \beta_{2N}^{L-1} \psi^3_{2N}
        \end{bmatrix}
        \begin{pmatrix}
            c_1 \\ c_2 \\ \vdots \\ c_{2N}
        \end{pmatrix}
        =
        \begin{pmatrix}
            0 \\ 0 \\ \vdots \\ 0
        \end{pmatrix}
        , \label{eq:det}
    \end{align}
    with some vectors $\varphi^1_l$, $\varphi^2_l$, $\varphi^3_l$, $\psi^1_l$, $\psi^2_l$, and $\psi^3_l$ ($l=1,\ldots,2N$), which are independent of $L$. The linear system \cref{eq:eq-finite} with $F=0$ admits a nontrivial solution when the determinant of the $2N\times 2N$ matrix in \cref{eq:det} vanishes. This is an algebraic equation in terms of $\beta_1,\ldots,\beta_{2N}$. By the argument in the Supplemental Material of \cite{yokomizo2019non-bloch}, the two values $\beta_N$ and $\beta_{N+1}$ must have the same modulus in order that they form a continuum band as $L\to\infty$.
}

{
    The spectrum $\sigma_\mathrm{GBZ}$ is numerically identified based on the following algorithm. For various $\omega\in\mathbb C$, we solve the eigenvalue problem \cref{eq:periodic-beta} in terms of $\beta$ and sort the eigenvalues $\beta_1(\omega),\ldots,\beta_{2N}(\omega)$ according to their modulus. If the difference $||\beta_N(\omega)|-|\beta_{N+1}(\omega)||$ is sufficiently small, the angular quasifrequency $\omega\in\mathbb C$ is regarded as a member of $\sigma_\mathrm{GBZ}$.
}

\begin{figure}
    \centering
    \includegraphics[scale=0.6]{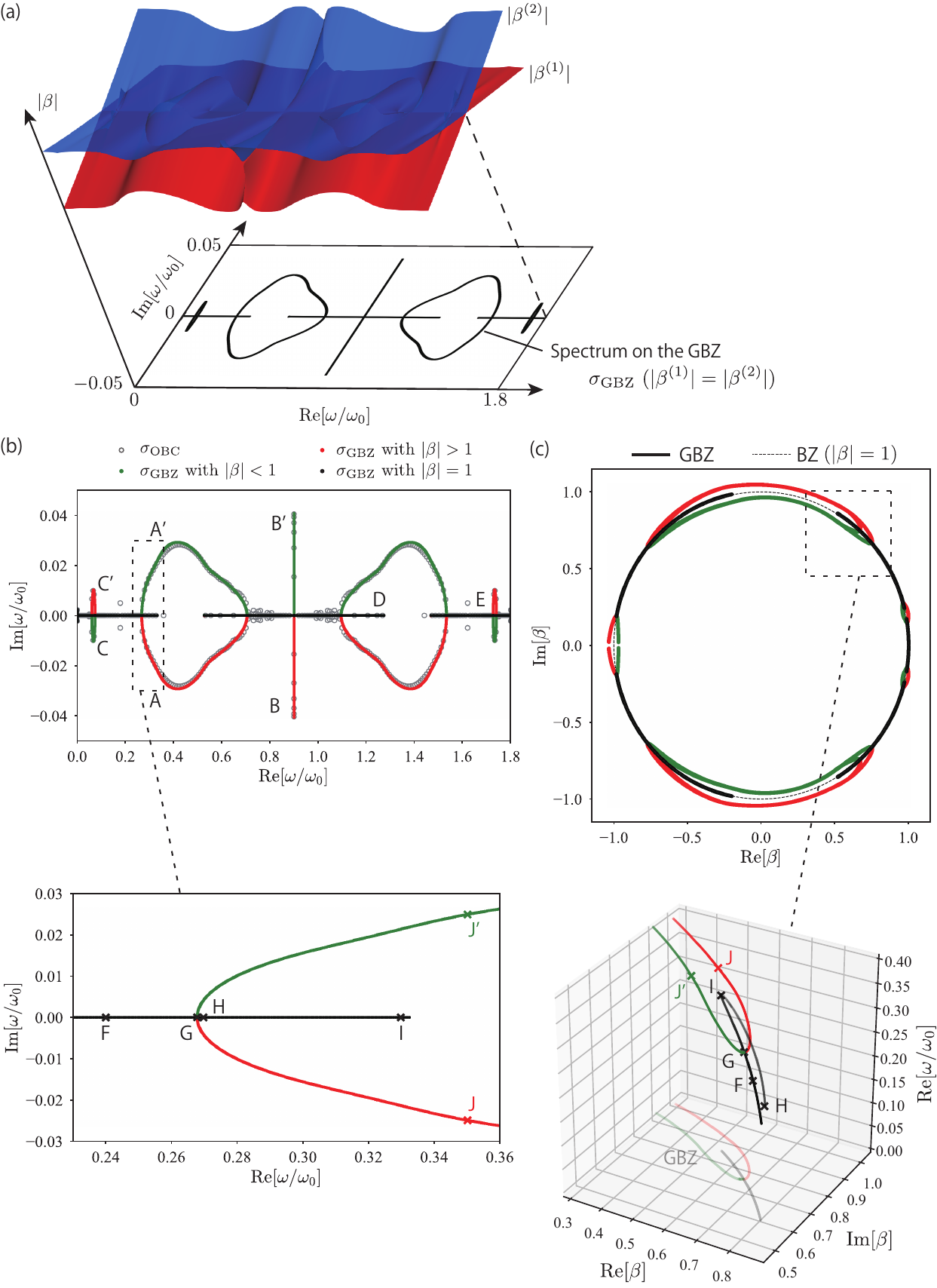}
    \caption{Non-Bloch band theory for the periodic chain shown in \cref{fig:result-periodic-spectrum_result} (a). (a) Eigenvalues $\beta(\omega)$ of \cref{eq:periodic-beta} as a function of $\omega\in\mathbb{C}$. The contour $|\beta^{(1)}(\omega)|=|\beta^{(2)}(\omega)|$ is shown in the complex $\omega$-plane. (b) Comparison between the GBZ spectrum $\sigma_\mathrm{GBZ}$ and $\sigma_\mathrm{OBC}$ shown in \cref{fig:result-finite-spectrum_result}. (c) $C_\mathrm{GBZ}$ in the complex $\beta$-plane. The red, green, and black lines represent $|\beta|>1$, $|\beta|<1$, and $|\beta|=1$, respectively.}
    \label{fig:result-periodic-sweep-beta_result}
\end{figure}
We numerically show that this non-Bloch band theory predicts the spectrum of the finite-length time-modulated system. We use the same parameters and compute $\beta^{(1)} = \beta_{ N}$ and $\beta^{(2)} = \beta_{ N+1}$ for varying $\omega\in\mathbb{C}$. The results are plotted in \cref{fig:result-periodic-sweep-beta_result} (a) and indicate that the contour $|\beta^{(1)}(\omega)| = |\beta^{(2)}(\omega)|$ (i.e. $\sigma_\mathrm{GBZ}$) forms a continuous curve with several branches in the complex $\omega$-plane, which resembles the finite-chain spectrum $\sigma_\mathrm{OBC}$ in \cref{fig:result-finite-spectrum_result} (c). In fact, \cref{fig:result-periodic-sweep-beta_result} (b) shows that almost all the eigenvalues in $\sigma_\mathrm{OBC}$ lie on the contour $\sigma_\mathrm{GBZ}$. \cref{fig:result-periodic-sweep-beta_result} (d) shows $C_\mathrm{GBZ}$ calculated from the spectrum $\sigma_\mathrm{GBZ}$. In contrast to the standard BZ (unit circle $|\beta|=1$), the GBZ deviates from the circle, which suggests the existence of non-Hermitian skin effects \cite{yao2018edge}. An eigenmode corresponding to an eigenvalue $\omega\in\sigma_\mathrm{GBZ}$ is called a non-Hermitian skin mode if $|\beta_N(\omega)| = |\beta_{N+1}(\omega)| \neq 1$. For example, the eigenvalue A in \cref{fig:result-periodic-sweep-beta_result} (b) lies on the spectrum $\sigma_\mathrm{GBZ}$ with $|\beta|>1$. This implies that the corresponding eigenmode grows exponentially as the site index $j$ increases. In fact, the finite-length chain exhibits a growing eigenmode at A as we have seen in \cref{fig:result-finite-spectrum_result} (d). The same discussion is valid for other eigenvalues B, C, D, E, and their conjugate pairs, i.e., the non-Bloch band theory with the GBZ tells whether a given eigenvalue is associated with a mode localized at the edges (non-Hermitian skin mode) or extending in the bulk (usual Bloch mode).

Time modulation gives rise to non-Hermitian degeneracy of eigenvalues. In the zoom-in views in \cref{fig:result-periodic-sweep-beta_result} (b) and (c), we mark some representative points F, G, H, I, and J. When the eigenvalues $\omega\in\sigma_\mathrm{GBZ}$ are seen as a function of $\beta$, we can observe that two eigenvalues $\omega$ degenerate at around the point I. Numerical experiments show that the matrix $\mathcal{H}_k$ has two degenerate eigenvalues $\omega/\omega_0 = 0.32998$ at $k=0.99805$ ($\beta=0.54194+0.84042\mathrm{i}$). We calculated the $\ell^2$ inner product between the corresponding eigenvectors $\varphi_1$ and $\varphi_2$ and obtained $|(\varphi_1/\| \varphi_1\|_{\ell^2},\varphi_2/\| \varphi_2\|_{\ell^2})_{\ell^2}| = 0.0974$, which implies that the eigenvectors are non-orthogonal in the $\ell^2$ sense. This cannot be observed in non-modulated systems as the matrix (operator) $\mathcal{H}_k$ is Hermitian (self-adjoint in $\ell^2$). 

{
    The existence of non-Hermitian skin modes is closely related to the reciprocity of a system. The periodic chain with three mass-spring pairs is spatially asymmetric due to the phase differences of stiffness modulation. This asymmetry is necessary for breaking reciprocity. Reciprocity implies that eigenfrequencies $\omega$ are invariant with respect to the wavenumber inversion $k\to-k$, meaning that each eigenfrequency trajectory does not form a closed loop encircling a nonzero area when $k$ varies along the BZ (from $-\pi$ to $\pi$). This concludes that the non-Hermitian skin effect does not manifest itself \cite{lin2023topological}. In fact, when the modulation phase $\phi_3$ is changed as $\phi_3 = 0$, i.e., $\phi_1=\phi_3$, the system is no longer non-reciprocal \cite{ammari2022nonreciprocal}. In this case, we numerically observed that the GBZ coincides with the standard BZ (unit circle $|\beta|=1$ in the complex $\beta$-plane) while the corresponding quasifrequency spectrum comprises complex numbers with positive and negative imaginary parts. From this observation, we conclude that both the spatial asymmetry and stiffness modulation are essential for achieving the non-Hermitian skin effect, characterized by the discrepancy between the standard BZ and GBZ. In such cases, the proposed GBZ theory gives useful and quantitative information about non-Hermitian skin modes.
}

\subsection{Transient wave motion}
We finally discuss the transient wave motion in the time-modulated system. As shown in \cref{fig:result-finite-spectrum_result} (a), we consider the finite-length chain with $M=300$ and the same parameters in \cref{ss:example}. 

Here, we numerically solve the system of ordinary differential equations \cref{eq:tmp1,eq:tmp2,eq:tmp3} using the backward Euler method with time increment $T/1000$, as we have done in \cref{ss:example}. The initial condition is set to $u^j=0$ and $\partial_t u^j=0$ for all $j=1,\ldots,M$. A time-harmonic external force is applied at the center of the chain with $M=300$, i.e., 
\begin{align}
    f_j(t) = 
    \begin{cases}
        A \cos \omega_\mathrm{f} t & (j=151)
        \\
        0 & (\text{otherwise})
    \end{cases}
\end{align}
with the angular frequency $\omega_\mathrm{f} = 0.31\omega_0$ and constant amplitude $A>0$. {Note that the phase of the forcing does not affect the amplitudes of excited modes.}

\begin{figure}
    \centering
    \includegraphics[scale=0.8]{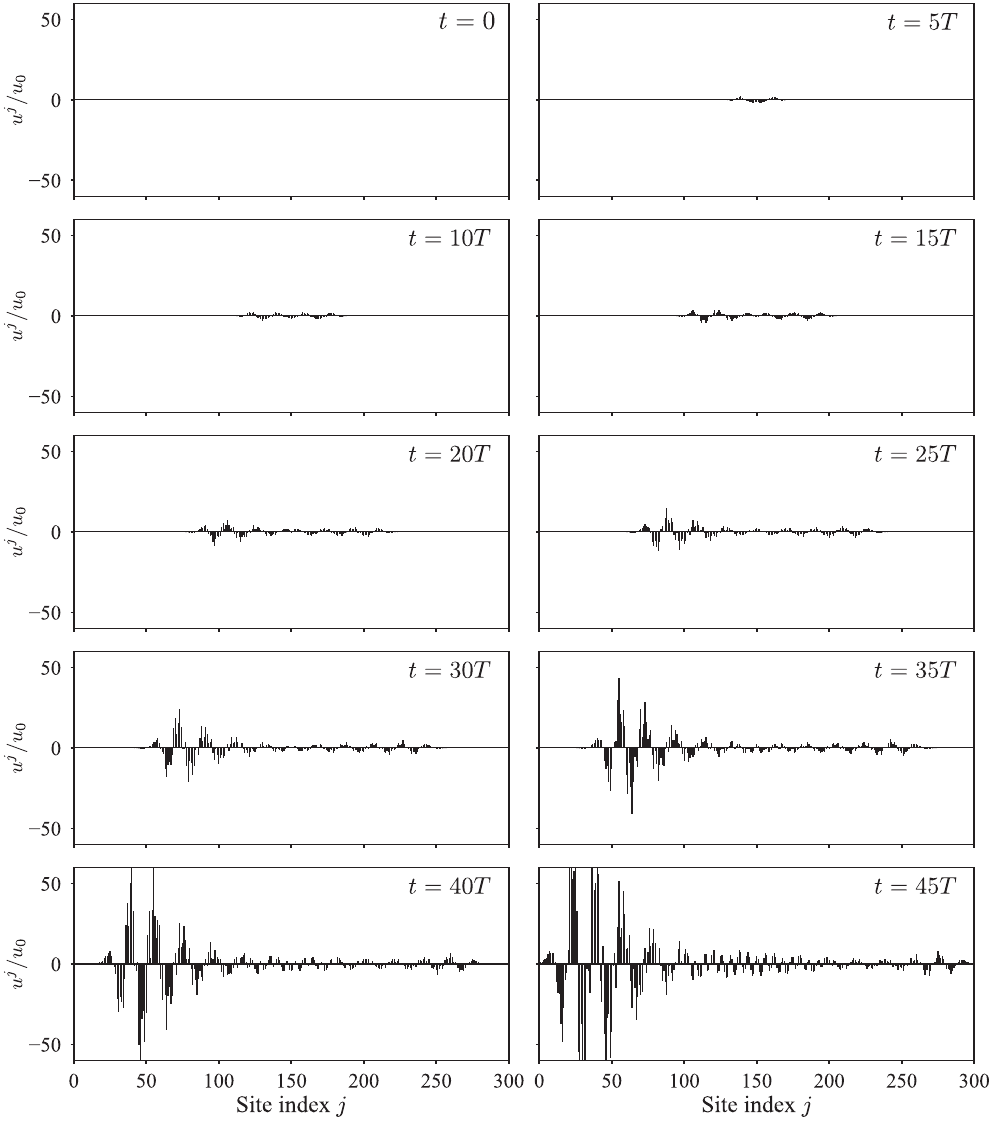}
    \caption{Transient wave motion in the finite-length chain shown in \cref{fig:result-finite-spectrum_result} (a) at some time steps. The displacements $u^j$ are normalized by $u_0 := A/G$.}
    \label{fig:result-time_result}
\end{figure}
\cref{fig:result-time_result} shows the displacements $u^j$ at some time steps. The mechanical waves excited by the external force propagate both in the left and right directions. While the right-propagating wave almost preserves its amplitudes, the left-propagating wave rapidly grows in time before reaching the edge. This asymmetric (non-reciprocal) propagation can be explained using the non-Bloch band theory. \cref{tab:result-time_result} lists some eigenvalues $\beta$ of \cref{eq:periodic-beta} and corresponding wavenumber $k$ for $\omega=\omega_\mathrm{f}$. The table indicates that the system has only two eigenvalues $\beta$ with $|\beta|=1$ (or equivalently $\mathrm{Im}[k]=0$). As both the two eigenvalues $\beta_N$ and $\beta_{N+1}$ has positive imaginary part (positive real part in $k_{N}$ and $k_{N+1}$), the array supports no propagating waves with constant amplitude (Bloch modes) in the left direction. The other eigenvalues $\beta_{N-1}$ and $\beta_{N+2}$ stand for left-propagating waves with growing and decaying in space (non-Hermitian skin modes), respectively. This is consistent with the transient wave motion in \cref{fig:result-time_result}.

\begin{table}[]
\caption{Eigenvalues $\beta$ of \cref{eq:periodic-beta} at $\omega = \omega_\mathrm{f}$.}
\begin{tabular}{c|cc}
$r$  & $\beta_r$ & $k_r=-\mathrm{i}\log \beta_r$ \\ \hline 
$N-1$ & $5.564 \times 10^{-1} -6.880 \times 10^{-1}\mathrm{i}$ & $-8.908 \times 10^{-1} + 1.223 \times 10^{-1}\mathrm{i}$  \\
$N$ & $5.748 \times 10^{-1} + 8.183 \times 10^{-1}\mathrm{i}$ & $9.584 \times 10^{-1} + 2.656 \times 10^{-14}\mathrm{i}$  \\
$N+1$ & $6.850 \times 10^{-1} + 7.286 \times 10^{-1}\mathrm{i}$ & $8.163 \times 10^{-1} -2.504 \times 10^{-14}\mathrm{i}$  \\
$N+2$ & $7.106 \times 10^{-1} -8.787 \times 10^{-1}\mathrm{i}$ & $-8.908 \times 10^{-1} -1.223 \times 10^{-1}\mathrm{i}$  \\
\end{tabular}
\label{tab:result-time_result}
\end{table}

\section{Concluding remarks}
In this study, we proposed a non-Bloch band theory for estimating eigenvalue distribution of time-modulated mechanical systems. The temporal Floquet theory revealed an analogy with non-Hermitian physics in quantum mechanics and suggested the breakdown of the conventional Bloch band theory. We presented some numerical examples and observed significant discrepancy between the spectra of finite and periodic systems due to the presence of non-Hermitian skin effects. To deal with such localized modes, we proposed a non-Bloch band theory based on a GBZ. Numerical experiments showed that the results from the proposed theory are consistent with the spectrum for a finite-length chain and transient wave motion calculated using the backward Euler scheme. 

{
    As we focused on eigenvalues and eigenmodes of Floquet systems without forcing, a future direction of this work is a more detailed analysis of the excitation of Bloch or non-Hermitian skin modes in response to given forces. For example, we are interested in the relationship between excited amplitudes and modulation or forcing phases, which is well studied in single resonator models described by the Mathieu equation \cite{acar2016floquet}. In addition, higher-dimensional lattices are of interest as such systems can possibly exhibit edge/corner skin effects or unidirectional mode conversion.
}

{
    We note that other mathematical tools may also be useful for the analysis of time-modulated systems. For instance, a higher-order homogenization method for the wave equation with time-varying coefficients was recently proposed in \cite{touboul2024high}. It is worth discussing whether such homogenization-based approaches are valid for time-modulated and non-reciprocal models exhibiting the non-Hermitian skin effect.
}


\section*{Acknowledgements}
This work was supported by JSPS KAKENHI Grant Numbers 24K17191, 23H03413, 23H03798. KM acknowledges support from Mizuho Foundation for the Promotion of Sciences.

 





\end{document}